\documentstyle[amssymb,preprint,aps,epsf,floats]{revtex}

\tighten
\preprint{\vbox{
\hbox{DOE/ER/40762-254}
\hbox{UMD-PP-02-049}
}} \bigskip \bigskip

\begin{document}
\title{Connecting the Quenched and Unquenched Worlds via the Large N$_{c}$ World}
\author{Jiunn-Wei Chen}
\address{Department of Physics, University of Maryland, \\
College Park, MD 20742, USA\\
{\tt jwchen@physics.umd.edu}}
\maketitle

\begin{abstract}
In the large $N_{c}$(number of colors) limit, quenched QCD and QCD are
identical. This implies that, in the effective field theory framework, some
of the low energy constants in ($N_{c}=3$) quenched QCD and QCD are the same
up to higher-order corrections in the 1/$N_{c}$ expansion. Thus the
calculation of the nonleptonic kaon decays relevant for the $\Delta I=1/2$
rule in the quenched approximation is expected to differ from the unquenched
one by an ${\cal O}$(1/$N_{c}$) correction. However, the calculation
relevant to the CP-violation parameter $\epsilon ^{\prime }/\epsilon $ would
have a relatively big higher-order correction due to the large cancellation
in the leading order. Some important weak matrix elements are poorly known
that even constraints with 100\% errors are interesting. In those cases,
quenched calculations will be very useful.
\end{abstract}

\vfill\eject

Quantum chromodynamics (QCD) is the underlying theory of strong interaction.
At high energies ($\gg 1$ GeV), the consequences of QCD can be studied in
systematic, perturbative expansions. Good agreement with experiments is
found in the electron-hadron deep inelastic scattering and hadron-hadron
inelastic scattering (Drell-Yan process) \cite{PQCD}. At low energies ($%
\lesssim 1$ GeV), however, QCD becomes non-perturbative. Quantitative
results not based on symmetry properties along can only be calculated
directly from QCD via computer simulations on a Euclidean lattice.

In lattice QCD\ simulations, it is common to perform the quenched
approximation which is to drop the internal quark loop contributions by
setting the fermion determinant arising from the path integral to be one.
This approximation cuts down the computing time tremendously and by
construction, the results obtained are close to the unquenched ones when the
quark masses are heavy ($\gg \Lambda _{QCD}$ $\sim 200$ MeV) so that the
internal quark loops are suppressed. Thus the quenched approximation is an
efficient and reliable tool to use as long as all the quark masses are
heavy. However, in the real world, the light quark ($u,d$ and $s$) masses
are less than $\Lambda _{QCD}$ so the justification for quenching does not
apply. In this range of quark mass, quenching effects can be systematically
studied by quenched chiral perturbation theory (Q$\chi $PT) \cite{QChPT}, an
effective field theory of quenched QCD (QQCD). In Q$\chi $PT, quenching
could induce a double pole to the flavor singlet meson propagator and could
make quenched quantities more singular than the unquenched ones \cite{QChPT}
in the infrared region. In the ultraviolet region, QCD and QQCD are
different as well because QQCD can be considered as having infinite quark
masses for the internal loops so its ultraviolet behavior is different from
that of QCD. Those uncontrolled effects make it hard to see what we can
learn about QCD from QQCD.

However, there is a known connection between QCD and QQCD in the limit that
the number of colors ($N_{c}$) becomes infinite; the large $N_{c}$ limit of
QQCD is indistinguishable to the large $N_{c}$ limit of QCD. In this paper,
we study the finite $N_{c}$ quenching errors systematically using effective
field theories with $1/N_{c}$ expansions. Generally, in the real world where 
$N_{c}=3$, some of the contributions which are formally higher order in the $%
1/N_{c}$ expansion could become numerically important for certain
observables near the chiral limit (zero quark mass limit). Fortunately,
these big contributions are typically non-analytic in quark masses and can
be computed in effective field theories of QCD and QQCD. After these big
non-analytic contributions in QQCD are replaced by those in QCD, it is
plausible that the errors which are formally $O(1/N_{c})$ in the quenched
approximation could imply numerically $\sim 30\%$ errors. Sometimes even
smaller errors can be achieved by forming ratios of observables, such that
the $O(1/N_{c})$ errors from the leading terms in the chiral expansion
cancel. The ratio of $B$ meson decay constants $f_{B_{s}}/f_{B}$ \cite
{SZ,fbratio} is an explicit example. We then study the quenched calculations
of the non-leptonic kaon decays $K\rightarrow \pi \pi $\cite{kaon1,kaon2}.
We find that the quenching error in the part relevant to the $\Delta I=1/2$
rule is consistent with $O(1/N_{c})\sim 30\%.$ However, the CP violation
parameter $\epsilon ^{\prime }/\epsilon $ calculation could have a big
quenching error due to the large cancellation in the leading terms in the $%
1/N_{c}$ expansion. Finally, for illustration, we identify some cases that
quenched calculations can be very useful. Those are calculations of
important but poorly known weak matrix elements for which even $100\%$ error
constraints are interesting.

\begin{figure}[t]
\begin{center}
\epsfxsize=10.25cm
\centerline{\epsffile{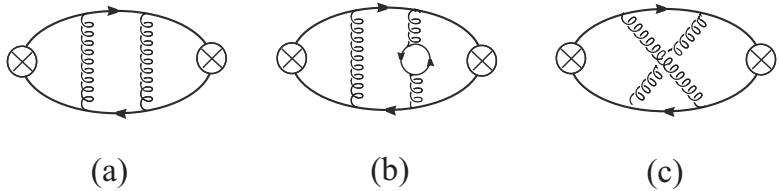}}
\end{center}
\caption{Feynman diagrams contributing to the pion two point function. The
solid lines and curly lines denote the quark and gluon propagators
respectively.}
\end{figure}

It is known that the quenched world has some similarity to the large $N_{c}$
world. Take a pion two-point function $\left\langle 0\left| \pi (x)\pi
(0)\right| 0\right\rangle $, for example: there are contributions from
diagrams shown in Fig. 1. For convenience, the quarks connected to the
external sources are called ``valence quarks''; the quarks not connected to
the external sources are called ``sea quarks''. In the quenched
approximation, the diagrams with sea quark loops such as the one in Fig.
1(b) are zero. (Then the gluon-quark coupling in surviving diagrams such as
the ones in Fig. 1(a,c) are usually rescaled to mimic the situation that the
sea quark masses are set to infinity. We will not do this rescaling at this
moment but leave it for later discussion.) In contrast, in the large $N_{c}$
limit proposed by 't Hooft \cite{thooft}, 
\begin{equation}
N_{c}\rightarrow \infty \text{, \ }g\rightarrow 0\text{, \ }%
N_{c}g^{2}\rightarrow \text{constant,}  \label{one}
\end{equation}
where $g$ is the quark gluon coupling, both the Fig.1(b) diagram with one
sea quark loop and the ``non-planar'' diagram of Fig.1(c) with one crossing
in gluon lines are suppressed by one power of $1/N_{c}$ compared to that of
Fig.1(a). The diagrams with more sea quark loops or more gluon crossings are
further suppressed by even higher powers of $1/N_{c}$. So, in the large $%
N_{c}$ limit of QCD, only the planar diagrams such as Fig.1(a) survive,
while in QQCD both Figs.1(a) and (c) survive. Now if we further take the
large $N_{c}$ limit of QQCD, the non-planar diagrams vanish as well. From
this we conclude that QQCD and QCD have the same large $N_{c}$ limit---at
least in the mesonic two-point function case. It is easy to generalize the
above observation to an arbitrary n-point function even involving baryons.
To generalize the baryons to the large $N_{c}$ case, we follow the
prescription suggested by Dashen and Manohar \cite{DM} such that there is no
valence strange quark in a proton.

The identity between the large $N_{c}$ QCD and large $N_{c}$ QQCD imposes
some constraints on the low energy effective field theories of QCD and QQCD,
which are the chiral perturbation theory ($\chi $PT) \cite
{ChPT,GL,ChPTJM,ChPTUlf} and the quenched chiral perturbation theory (Q$\chi 
$PT) \cite{QChPT}, respectively. $\chi $PT and Q$\chi $PT have the same
symmetries and symmetry breaking patterns as QCD and QQCD. In both cases,
low energy observables are expanded in a power series of the small expansion
parameters $(p,M_{GB})/\Lambda _{\chi }$, where $p$ is the characteristic
momentum transfer in the problem, $M_{GB}$ is the goldstone boson mass and $%
\Lambda _{\chi }$ is an induced chiral perturbation scale. In QCD, $\Lambda
_{\chi }\sim 1$ GeV and the goldstone boson masses are $M_{\pi }\sim 140$
MeV, $M_{K}\sim 500$ MeV and $M_{\eta }\sim 550$ MeV. In the QCD case ($%
N_{c}=3$ ), $\chi $PT gives \cite{GL} 
\begin{eqnarray}
M_{\pi }^{2} &=&2m_{u}B\left\{ 1+\frac{1}{(4\pi f_{\pi })^{2}}\left[ M_{\pi
}^{2}\log \left( \frac{M_{\pi }^{2}}{\mu ^{2}}\right) -\frac{1}{3}M_{\eta
}^{2}\log \left( \frac{M_{\eta }^{2}}{\mu ^{2}}\right) \right] \right. 
\nonumber \\
&&+\left. 2m_{u}K_{3}+\frac{2m_{u}+m_{s}}{3}K_{4}\right\} +O(m_{q}^{3})\ ,
\label{mpiqcd}
\end{eqnarray}
where $m_{q}$ is the mass of the flavor $q(=u,d,s)$ quark and we have used $%
m_{d}=m_{u}$. The pion decay constant $f_{\pi }=93$ MeV and the
renormalization scale $\mu $ dependence is absorbed by the low energy
constants (or counterterms) $K_{3}$ and $K_{4}$ defined in \cite{GL}. When $%
N_{c}$ becomes large, $\ f_{\pi }={\cal O}(\sqrt{N_{c}})$, $K_{4}={\cal O}%
(1/N_{c}),$ while $B$, quark masses and $K_{3}$ are ${\cal O}(1)$. Thus as $%
N_{c}\rightarrow \infty $, only the $B$ and $K_{3}$ terms contribute. 
\begin{equation}
M_{\pi }^{2}\rightarrow 2m_{u}B\left( 1+2m_{u}K_{3}\right) +O(m_{q}^{3}).
\label{LN1}
\end{equation}
Note that the $\mu $ dependence of $K_{3}$ is ${\cal O}(1/N_{c})$, so $K_{3}$
becomes $\mu $ independent as $N_{c}\rightarrow \infty $. In the real world,
the chiral logarithms could be numerically enhanced even though they are
formally higher order in $1/N_{c}$.

In the QQCD ($N_{c}=3$) case, Q$\chi $PT gives a similar result for the pion
mass \cite{QChPT} 
\begin{equation}
M_{\pi }^{Q^{2}}=2m_{u}B^{Q}\left\{ 1-\frac{2}{3(4\pi f_{\pi }^{Q})^{2}}%
M_{0}^{2}\log \left( \frac{M_{\pi }^{Q^{2}}}{\mu ^{2}}\right)
+2m_{u}K_{3}^{Q}\right\} +\cdots ,  \label{qmpi}
\end{equation}
where the low energy constants $B^{Q}$, $\ f_{\pi }^{Q}$ and $K_{3}^{Q}$
have different values than those in QCD and where $M_{0}^{2}$ is the
strength of the $\eta ^{\prime }\eta ^{\prime }$ coupling. In QCD,\ $%
M_{0}^{2}$ can be iterated to all orders to develop a shifted $\eta ^{\prime
}$ pole such that $\eta ^{\prime }$ becomes non-degenerate with the
goldstone bosons in the chiral limit. However, in QQCD, all the sea quark
loops are dropped such that the iteration stops at one insertion of $%
M_{0}^{2}$. This gives the $\eta ^{\prime }$ propagator a double pole
dependence and makes physical quantities in QQCD more singular than those in
QCD in the infrared region\cite{QChPT}. The quenched pion mass in eq.(\ref
{qmpi}) is an explicit example. The quenched chiral logarithm (the $%
M_{0}^{2} $ term) is more singular than the QCD chiral logarithm in eq.(\ref
{mpiqcd}). As $N_{c}$ becomes large, $M_{0}^{2}$ scales as $1/N_{c}$ while
the other quenched low energy constants scale the same way as the
corresponding unquenched ones, $\ f_{\pi }^{Q}$ $={\cal O}(\sqrt{N_{c}})$, $%
B^{Q}$ and $K_{3}^{Q}$ $={\cal O}(1).$ Thus as $N_{c}\rightarrow \infty $, 
\begin{equation}
M_{\pi }^{Q^{2}}\rightarrow 2m_{u}B^{Q}\left( 1+2m_{u}K_{3}^{Q}\right)
+O(m_{q}^{3}).  \label{LN2}
\end{equation}

Since QCD and QQCD have the same large $N_{c}$ limits, eqs.(\ref{LN1}) and (%
\ref{LN2}) imply 
\begin{equation}
B=B^{Q}\text{ , }K_{3}=K_{3}^{Q}\text{ as }N_{c}\rightarrow \infty \text{.}
\end{equation}
Thus some low energy constants (to be more precise, those that are leading
in the $1/N_{c}$ expansion)\ of $\chi $PT and Q$\chi $PT are identical in
the large $N_{c}$ limit. This implies their differences are higher order in $%
1/N_{c}$: 
\begin{eqnarray}
B^{Q} &=&B\left( 1+{\cal O}(\frac{1}{N_{c}})\right) ,  \nonumber \\
K_{3}^{Q} &=&K_{3}\left( 1+{\cal O}(\frac{1}{N_{c}})\right) .  \label{LEC1}
\end{eqnarray}

Now we can have a clear idea about how large the quenching error is. The
error could come from the quenched logarithms which might be formally higher
order in the $1/N_{c}$ expansion but numerically significant. Fortunately,
those chiral logarithms can be calculated using effective field theories so
that one can just replace the quenched logarithms by the real ones. Another
source of error comes from the difference between the low energy constants
which is an ${\cal O}(1/N_{c})\sim 30\%$ effect. This is consistent with the 
$\sim 25\%$ quenching error seen in simulations (see \cite{fbratio}, for
example).

Recently kaon weak matrix elements relevant to the $K\rightarrow \pi \pi $
process have been simulated\ using the quenched approximation by two groups 
\cite{kaon1,kaon2}. In both calculations, Q$\chi $PT low energy constants
are extracted from $K\rightarrow \pi $ and $K\rightarrow vaccum$ amplitudes
in order to recover the desired $K\rightarrow \pi \pi $ amplitude. Based on
the above discussion that leads to eq.(\ref{LEC1}), the errors of these
analyses are ${\cal O}(1/N_{c})$ and naively are $\sim 30\%$. Indeed this is
the case for the $\Delta I=1/2$ rule. The simulations give $25.3\pm 1.8$ 
\cite{kaon1} and $10\pm 4$\ \cite{kaon2} compared to the experimental value $%
22.2$. However, for the case of CP violation $\epsilon ^{\prime }/\epsilon $
parameter, Ref. \cite{kaon1} obtained $(-4.0\pm 2.3)\times 10^{-4}$,
consistent with Ref. \cite{kaon2}, compared to the current experimental
average of $(17.2\pm 1.8)\times 10^{-4}$. In the $\epsilon ^{\prime
}/\epsilon $ case, a large accidental cancellation happens between the $I=0$
and $I=2$ contributions such that the estimated $\sim 30\%$ errors from the
two pieces add up to $\sim 8\times 10^{-4}$. Besides, in the $\epsilon
^{\prime }/\epsilon $ quenched calculation, there is an additional quenched
artifact that can be fixed by dropping some ``eye diagrams'' \cite{GP}.
Doing this will increase the $\epsilon ^{\prime }/\epsilon $ by $\sim
4\times 10^{-4}$ \cite{Blum}. Thus in this case the quenching error is
significant.

The above examples suggest that in principle, the quenched approximation is
typically expected to give a $\sim 30\%$ error result if there are no big
cancellations between different contributions and if the quenched chiral
logarithms (or non-analytic contributions, in general) are replaced by the
unquenched ones. Now we will try to identify some matrix elements that can
be reliably computed in the quenched approximation with $100\%$ errors and
are still worth computing even with that size of uncertainties.

The first quantity we look at is the proton strange magnetic moment $\mu
_{s}.$ It is measured in the electron-proton and electron-deuteron parity
violation experiments \cite{SAMPLE}. The measured value has a large
uncertainty 
\begin{equation}
\mu _{s}^{\exp }=\left[ 0.01\pm 0.29(\text{stat})\pm 0.31(\text{sys})\pm
0.07(\text{theor.})\right] \text{ n.m.}
\end{equation}
This implies that the strange quark only contributes around $-0.1\pm 5.1\%$
of the proton's magnetic moment. In $\chi $PT, the first two orders in
chiral expansion for $\mu _{s}$ is \cite{musChPT}

\begin{equation}
\mu _{s}=\mu _{s}^{0}+\frac{M_{N}M_{K}}{24\pi f_{\pi }^{2}}%
(5D^{2}-6DF+9F^{2}),
\end{equation}
where $D$ and $F$ are goldstone boson nucleon couplings. Using the standard
values for the parameters, the second term which is the kaon cloud
contribution is $\sim 2.0$ n.m. This combined with the experimental data
implies the leading order low energy constant $\mu _{s}^{0}$ is $\sim -2.0$
n.m. Thus a quenched calculation would have a quenching error of $\sim 0.8$
n.m. Again, the large cancellation makes the quenching error relatively big
and, hence, the result is not reliable within a $100\%$ error.

The second quantity that is also interesting but even less well measured is
the longest range parity violating (PV) isovector pion nucleon coupling, $%
h_{\pi }$. For many years, serious attempts have been made to measure this
quantity in many-body systems where the PV effects are enhanced. However,
currently even its order of magnitude is still not well constrained (see 
\cite{Oers} for reviews) largely due to the theoretical uncertainties.
Few-nucleon experiments, such as the new $\overrightarrow{n}p\rightarrow
d\gamma $ experiment at LANSCE \cite{LANSCE}, are expected to have small
theoretical uncertainties and will set a tight constraint on $h_{\pi }$.
Also, recent studies in single nucleon systems in Compton scattering \cite
{PVComp} and pion productions \cite{PVpion} show that the near threshold $%
\overrightarrow{\gamma }p\rightarrow \pi ^{+}n$ process is both a
theoretically clean and experimentally feasible way to extract $h_{\pi }$%
\cite{PVpion}. Given the intensive interests and the level of difficulty
involved in the experiments, a lattice QCD determination of $h_{\pi }$ with
a $100\%$ error will be very valuable. In fact, issues such as chiral
extrapolation has been addressed in \cite{BS} in the context of partially
quenched approximations \cite{PQQCD} in which sea and valence quark masses
are different. Since one does not expect large cancellations between the
first few orders---as what happened in the $\mu _{s}$ case---we suggest that
a quenched simulation of $h_{\pi }$ is enough to give a $100\%$ error result.

Another interesting PV quantity is the longest range PV photon-nucleon-delta
($\gamma N\Delta $) coupling. Zhu et al. suggested a model, which was
inspired by the hyperon decays, to argue that there could be enhanced PV
mixing in the initial and final state wave functions such that the PV $%
\gamma N\Delta $ coupling is 25-100 times larger than its size from the
naive dimensional analysis \cite{Zhu}. If correct, this enhancement will
introduce an easily measurable PV signal in $\overrightarrow{\gamma }%
p\rightarrow \pi ^{+}n$ at the delta peak. To check if the large PV
enhancement due to the wave function mixing really takes place, one can
compute the PV mixing between $\Delta (\frac{3}{2}^{+})$ and $N(1520;\frac{3%
}{2}^{-})$ and between the nucleon$(\frac{1}{2}^{+})$ and $N(1535;\frac{1}{2}%
^{-})$ in QQCD. One expects the mixing between these states to be 5-20 times
larger than their sizes from the naive dimensional analysis according to the
model of Zhu et al.. This can be checked easily by a quenched calculation
with a $100\%$ error.

Recently, Young et al. \cite{Young} studied the lattice results of nucleon
and delta masses. They suggested the difference between the quenched and
unquenched cases is largely due to the one-loop self-energy contributions
calculated in a certain cut-off scheme, while the difference due to the low
energy constants seems to be smaller than a generic estimation of $\sim 30\%$%
. This might be because in the real simulations, only the mass ratios are
computed. The absolute mass scale is set by assuming zero quenching error
for a certain quantity, then the coupling constant and, equivalently, the
lattice spacing, is determined accordingly. It is conceivable that the ratio
of certain quantities can be computed more reliably because the leading $%
{\cal O}(1/N_{c})$ errors cancel. An explicit example is the ratio of $B$
meson decay constants $f_{B_{s}}/f_{B}$, where $B_{s}(B)$ denotes a
low-lying pseudoscalar meson with the same quantum number as a $\overline{b}%
s(\overline{b}d)$ state \cite{SZ}. However, to argue that all the mass
ratios have smaller than ${\cal O}(1/N_{c})$ quenching errors is beyond the
scape of effective field theory. To see if certain error suppression really
takes place, it would be very interesting to study the $N_{c}$ dependence
for quenched approximations by direct simulations.

There is another class of ratios in large $N_{c}$ QCD that the finite $N_{c}$
correction is ${\cal O}(1/N_{c}^{2})$ to the leading order. It would be
interesting to see whether the correction is still ${\cal O}(1/N_{c}^{2})$
in QQCD.

In summary, quenched QCD and QQCD have the same large $N_{c}$ limit, such
that some of the low energy constants in their corresponding effective field
theories are identical in the large $N_{c}$ limit. This implies that some
physical quantities can be determined in the quenched approximation with $%
{\cal O}(1/N_{c})\sim 30\%$ accuracy. A few interesting applications are
explored.

\acknowledgements
The author thanks Tom Cohen, Xiangdong Ji, Martin Savage and Steve Sharpe 
for useful discussions. This work is supported in part by the U.S. Dept. of
Energy under grant No. DE-FG02-93ER-40762.

\end{document}